\documentclass[pra,showpacs,twocolumn,longbibliography,secnumarabic]{revtex4-1}
\usepackage{graphicx}
\usepackage{amssymb}
\usepackage{float}
\usepackage[T1]{fontenc}
\usepackage{amsmath}
\begin{document}
\title
{Environmental-induced work extraction}
\author{Rasim Volga Ovali$^{\dagger,1}$}
\author{Shakir Ullah$^{\dagger,2}$}
\author{Mehmet G\"{u}nay$^{\dagger,3}$}
\author{Mehmet Emre Tasgin$^{2,*}$}
\affiliation{$^\dagger$ Contributed equally}
\affiliation{$^*$ correspondence: metasgin@hacettepe.edu.tr and metasgin@gmail.com}
\affiliation{ $^{1}$Department of Physics, Recep Tayyip Erdogan University, 53100, Rize, Turkey}
\affiliation{ $^{2}$Institute  of  Nuclear  Sciences, Hacettepe University, 06800 Ankara, Turkey}
\affiliation{ $^{3}$Department of Nanoscience and Nanotechnology, Faculty of Arts and Science, Mehmet Akif Ersoy University, 15030 Burdur, Turkey}
\begin{abstract}
A measurement can extract work from an entangled, e.g., two-mode system. Here, we inquire the extracted work when no intellectual creature, like an ancilla/daemon, is present. When the monitoring is carried out by the environmental modes, that is when no measurement-apparatus is present, the measurement-basis becomes the coherent states. This implies a Gaussian measurement with a fixed strength $\lambda=1$. For two-mode Gaussian states, extracted work is already independent from the measurement outcome. After the strength is also fixed, this makes nature assign a particular amount of work to a given entanglement degree. Extracted work becomes the entanglement-degree times the entire thermal energy at low temperatures ---e.g., room temperature for optical modes. Environment, nature itself, converts entanglement to an ordered, macroscopic, directional~(kinetic) energy from a disordered, microscopic, randomized thermal energy. And the converted amount is solely determined by the entanglement.
\end{abstract}
\maketitle

Quantum entanglement enables technologies which are not possible without them~\cite{acin2018quantum}. Measurements below the standard quantum limit~\cite{LigoNaturePhot2013,pezze2018quantum}, quantum enhanced imaging~\cite{Lugiato_JOptB_2002,casacio2021quantum,pirandola2018advances}, quantum radars~\cite{maccone2020quantum}, quantum teleportation~(QT)~\cite{bennett1993teleporting}, and quantum computation~\cite{wright2019benchmarking} are all enabled by entangled ---more generally nonclassical~\cite{dodonov2002nonclassical}--- states. Entanglement can also be utilized as a resource for quantum heat engines~\cite{bresque2021two,josefsson2020double,hewgill2018quantum,zhang2007four,rossnagel2014nanoscale,dillenschneider2009energetics,scully2003extracting} which makes them operate more efficiently compared to their classical counterparts~\cite{kieu2004second}. As an example, an ancilla can utilize entanglement for extracting a larger amount of work by maximizing the efficiency~\cite{francica2017daemonic}. It is quite intriguing that entanglement can even be directly transformed into work via measurements~\cite{brunelli2017detecting,cuzminschi2021extractable}. The energy can be extracted from a single heat bath~\cite{brunelli2017detecting,cuzminschi2021extractable}~\footnote{Work extraction via a measurement can be performed also from the internal energy of the measurement apparatus~\cite{elouard2018efficient,elouard2017extracting}. However, here we are interested in the conversion of the microscopic energy (heat) present in a single heat bath into directional mechanical energy~\cite{brunelli2017detecting,cuzminschi2021extractable}.}.  
 This phenomenon ---we focus here--- takes place as follows, e.g., in a two-mode entangled state.

{\it Work extraction as a measurement backaction.}--- Let us assume that one of the modes (mode $a$) belongs to an optical cavity which includes, for example, a free-to-move board or a piston inside the cavity~\cite{brunelli2017detecting,cuzminschi2021extractable}. The second ($b$) mode relies somewhere outside of the cavity and it is entangled with the $a$-mode. Both modes are in thermal equilibrium with the environment at temperature T. When a measurement is carried out on the $b$-mode (outside), entropy of the cavity ($a$) mode decreases to $S_V^{\rm (meas)}$  because of the measurement backaction. After the measurement, the state ($a$-mode) thermalizes back to equilibrium and assigns a higher entropy $S_V^{\rm (ther)}$~\footnote{The entropy after the rethermalization is equal to the one before the measurement. More accurately, $S_V^{\rm (ther)}$ is also the half of the entropy belonging to the two-mode state before the measurement.}. During the rethermalization with the heat bath the expansion of the $a$-mode pushes the board located inside the cavity~\cite{brunelli2017detecting}. An amount of $W=k_B T( S_V^{\rm (ther)}-S_V^{\rm (meas)} )$ work can be extracted from the single heat bath. That is, $W$ amount of thermal energy can be converted into ``directional'' kinetic energy~(KE) of the board. In case of maximum entanglement, the state of the $a$-mode is completely determined from the outcome of the $b$-mode and entropy of the $a$-mode vanishes, i.e., $S_V^{\rm (meas)}=0$. Thus, the extractable work becomes $k_B T S_V^{\rm (ther)}$, i.e., the complete internal energy. The amount of extracted work can be employed for witnessing/quantifying the entanglement~\cite{oppenheim2002thermodynamical,maruyama2005thermodynamical,ciampini2017experimental,maruyama2009colloquium}.

In general, the extracted work depends on the nature of the measurement and its outcome. Yet, interestingly, the state of the $a$-mode~(cavity field) turns out to be independent of the outcome of the $b$-mode as long as Gaussian states and measurements are concerned~\cite{fiuravsek2002gaussian,giedke2002characterization,fiuravsek2007gaussian,giorda2010gaussian}.  The state, into which the $a$-mode collapses, depends only on the strength~($\lambda$) of the Gaussian measurement/operation carried out on the $b$-mode~\footnote{It also depends on the rotation angle 
$\phi$ belonging to the Gaussian operation. But this will not be our concern as will become apparent in the following text.}. Thus, also the extracted work depends only on $\lambda$. Here, $\lambda \in [0,\infty]$ is the quadrature-noise belonging to the Gaussian operation~\cite{eisert2002distilling}. That is, if $\lambda$ (for a reason) assigns a fixed value, a given degree of entanglement extracts a particular amount of work. This is the phenomenon we explore here.

The observable that is measured in an experiment~(this can be, for example, number of photons) is determined by the quantum apparatus employed in the measurement of the $b$-mode. More explicitly, monitoring of the environment on the apparatus~(i.e., decoherence) destroys the superpositions among the natural pointer states~(the measurement-basis). This makes us observe one of the values in the measurement-basis. Refs.~\cite{zurek2003decoherence,schlosshauer2019quantum,zurek1982environment,unruh1989reduction,zurek1981pointer,zurek2009quantum} provide mathematical and numerical demonstrations of the monitoring process.

{\it The question.}--- Here we examine the following already-answered-question in the context of work extraction process. What happens if no measurement apparatus is present? In other words, what is the pointer (measurement) basis if no intellectual being, such as a human, a daemon, or an ancilla, is present? In this case, environmental monitoring sets the measurement-basis as the the coherent states~\cite{zurek1993coherent,gallis1996emergence,tegmark1994decoherence,wiseman1998maximally,zurek1994decoherence,zurek1995decoherence,paraoanu1999selection,zurek1993preferred}. Measurement strength is $\lambda=1$ for any one of the coherent states.

{\it Environmental monitoring.}--- At this stage, we better re-depict the picture of the system we have in mind in a more clear way. The cavity ($a$)~mode is entangled with the $b$-mode. It is worth noting that, in general, the $a$ and $b$ modes do not need to be in interaction~\cite{pan1998experimental}. The $b$-mode is monitored by the environment. Environment~(a collection of infinite number of modes) can monitor the $b$-mode only indirectly as two light beams do not interact directly. Monitoring can be performed over common interactions with masses of particles present around which induces an effective interaction between the environment and the $b$-mode~\footnote{Coupling of a cavity field/mode to the input/output (vacuum) modes already employs exactly the same mechanism~\cite{gardiner2004quantum}.}.  It is worth noting that pointer states are still coherent states, for instance, in case a harmonic chain~\cite{tegmark1994decoherence} is considered. Therefore, in total, environment monitors (measures) the $b$-mode in one of the coherent states. So, the measurement is a Gaussian one.

It is straightforward to realize that the measurement strength is fixed $\lambda=1$. Moreover, a rotated $R(\phi)$ form of the coherent state basis ---$R(\phi)$ is present in the most general form of a Gaussian measurement~\cite{fiuravsek2002gaussian,giedke2002characterization,fiuravsek2007gaussian,giorda2010gaussian}--- is also a coherent state basis. Putting things together, nature itself makes a Gaussian measurement on one of the two entangled modes. The measurement basis is composed of coherent states. As the basis possesses a fixed $\lambda$: a particular amount of work~(KE) becomes assigned to a given degree of entanglement as long as Gaussian states are concerned.  The thermodynamical energy, probabilistic and disordered in nature, is converted into an ordered~(mechanical) form of energy now belonging to the board~\cite{brunelli2017detecting,cuzminschi2021extractable}. This sets an observer-independent, nature-assigned, association between entanglement and directional/ordered energy.  We call this phenomenon {\it environment-induced work extraction}~(EIWE).

One can gain a better understanding by examining the phenomenon in low-$T$ limit ---such as room temperature for optical resonances~\footnote{or for a Swinger pair creation (mass-energy conversion) process taking place over a critical electric field.}. At this limit, a simple-looking analytical result can be obtained, because the $k_B T$ term, present in the $W$ formula, cancels with a $1/k_B T$ term appearing within the entropy difference~\footnote{We can tell that the result is $T$-independent except a factor  $\bar{n}= 1/(e^{\hbar\omega_a/k_B T}-1)  \to e^{-\hbar\omega_a/k_B T}$ which stands for the probability of finding one of the two modes in the excited state. $(\bar{n}\hbar\omega_a)$ is already the energy present in the $a$-mode.}.  Please, see Eqs.~(S11) and (S13) in the Supplementary Material~(SM)~\cite{SM}. 

The extracted work $W=\xi(r)\: (\bar{n}\hbar\omega_a)$ depends only on the degree of the entanglement $\xi(r)=[1-2/(1+\cosh 2r)]$ which runs from $0$ to $1$ as entanglement increases. $\bar{n}$ is the occupation of the $a$ (cavity) mode of resonance $\omega_a$. So, $(\bar{n}\hbar\omega_a)$ is already the ``entire'' thermodynamical~(it is probabilistic) energy present inside the cavity either before the measurement or after the rethermalization process. Here we use the von Neumann entropy $S_V$ in difference to Ref.~\cite{brunelli2017detecting} where R\'{e}yni entropy is employed. $r$ is the two-mode squeezing rate which is proportional to the time the entanglement device is kept open~\cite{ScullyZubairyBook}.

We present the derivations in the SM~\cite{SM}. In the rest of the paper, we aim to put this result into words in order to develop a physical understanding.

We observe that the extracted work~(KE of the board or piston) is: the degree of entanglement times ``all of the thermal energy'' present inside the cavity~\footnote{Here, we prefer to use the words ``all of the thermal energy'' at equilibrium, because this is the grand canonical mean energy which is (classical) probabilistic. The energy present inside the cavity after the rethermalization is also probabilistic. Conservation of energy tells us the following. If the energy realized inside the cavity after the rethermalization assigns one of the classical probable ones, the extracted work needs to be equal to that value.}. It gets closer to $(\bar{n}\hbar\omega_a)$ in the case of maximum (max) entanglement~\footnote{Please note that energy of a maximum entangled two-mode Gaussian state approaches to infinite. Here, we confine ourselves to a regime where squeezing rate $r \ll \hbar\omega_a/k_BT$. The latter is typically $\sim 100$ for optical modes at room temperature. See the discussion above Eq.~(S11) in the SM~\cite{SM}}. In other words, max entanglement converts the entire thermodynamical energy of the ($a$) photon mode into the kinetic~(directional) energy of the board/piston~\footnote{Please note that this statement is valid at any temperature.}. As we will see below, this is true also for other max entangled states, e.g., max two-mode entangled state $(|1,0\rangle+|0,1\rangle)/\sqrt{2}$ and symmetrization entanglement of identical particles~\cite{tasginSymmetrizationEnt}. That is, we crosscheck our EIWE result with other incidences.

What is peculiar to EIWE is that the work is extracted by itself. That is, an observer-independent entanglement-energy correspondence appears.  The converted energy is proportional to the degree of the entanglement $\xi$~\footnote{It should be noted that one can quantify entanglement in various alternative ways such as entropy of the reduced state~\cite{entanglementEntropy},  symplectic spectrum~\cite{AdessoPRA2004} or via nonclassical depth~\cite{lee1991measure} when single-mode nonclassicality is wiped out~\cite{tasgin2020quantifications}. They, all, display parallel behavior to the measure logarithmic negativity $E_{\cal N}$ for Gaussian states~\cite{PlenioPRL2005LogNeg}. } and depends only on the excitation spectrum.

Before making the comparison with other systems, we would like to bring two important issues into attention. First, we note that $W$ is calculated using a density matrix which involves classical~(thermodynamical) probabilities ---grand canonical ensemble. It is a result weighted over classical probabilities. For this reason, we prefer to use the words ``all of the thermal energy'' present inside the cavity is converted into the KE of the board/piston. The energy present inside the cavity after the rethermalization is also probabilistic. Conservation of energy, however, tells us the following. If the energy realized inside the cavity after the rethermalization assigns one of the classically probable ones, the extracted work needs to be equal to that value. Second, we note that almost all of the notion~(e.g., entanglement-work conversion and entanglement-energy analogy~\cite{horodecki2001balance}) and the calculations carried out here are already discussed in other studies~\cite{brunelli2017detecting,cuzminschi2021extractable}. Here, in difference, we introduce the notion of entanglement-work correspondence due to the presence of nature-originated measurement.

{\it Comparison with other work extraction phenomena.}---  We first compare the (i)~EIWE result with the one for another (max) entangled state (ii)~$|e\rangle=(|1,0\rangle+|0,1\rangle)/\sqrt{2}$ in thermal equilibrium $\hat{\rho}=P(|0,0\rangle\langle0,0| + e^{-\hbar\omega_a/k_BT} |e\rangle\langle e|)$~\footnote{We ignore other excited states, such as $|0,1\rangle$ and $|1,0\rangle$, which actually have the same energy. We do this for the sake of a reasonable comparison only.}. When one measures the number of photons and the outcome  the $b$-mode is $|1\rangle$, $W=x\hbar\omega_a$ work is extracted in the cavity $a$-mode. Here, $x=e^{-\hbar\omega_a/k_BT}$ is the classical probability for realizing the two-mode system in the excited state $|e\rangle$ and it is equal to the occupation $\bar{n}$ at low $T$, i.e., $\bar{n}=x$. This result is the same with the max entanglement ($\xi=1$) case of EIWE. In this example, too, all thermodynamical energy present in the $a$-mode is converted into directional energy (work). In this case, however, the work extraction~(the same amount) is subject to the measurement of the $b$-mode in the excited state. In EIWE, in difference, any measurement outcome extracts this amount of work.

We also examine the work associated to the (iii)~symmetrization (max) entanglement~\cite{tasginSymmetrizationEnt} as a third example. In a recent study, we investigate the work extracted by identical particles in a system of $N$ total number of symmetrized bosons. The extracted work by $(N-1)$ particles is calculated when one of the (random) particles is measured in the excited state, of energy $\omega_{eg}$~\footnote{One should note that measuring the quantum state of a single particle in a condensate of $N$ identical particles is not a straightforward process. It necessitates certain rules/conditions which is studied experimentally~\cite{stamper1999excitation,stenger1999bragg,andersen2006quantized} and theoretically~\cite{tasgin2017many,tacsgin2011creation}. Please see Ref.~\cite{tasginSymmetrizationEnt} for a detailed analysis.}. In parallel with the previous cases, (i) and (ii), the extracted work, by pushing the board located within the condensate region, comes out as $W_3=x\hbar\omega_{eg}$ at low $T$. Here, $x=e^{-\hbar\omega_{eg}/k_B T}$ is the probability for one of the $N$ particles to occupy the excited state either before the measurement or after the rethermalization of the condensate. Similarly, $(x\hbar\omega_{eg})$ is the entire thermal energy of the condensate at equilibrium. The lowermost excited state of such a condensate is the Dicke state $|N,1\rangle$~\cite{mandel1995optical}, where a single-particle excitation is symmetrically distributed among $N$ bosons, $|N,1\rangle=(|e,g,g,...\rangle+|g,e,g,...\rangle...+|g,...g,e\rangle)/\sqrt{N}$. This is a maximally entangled state with respect to any one of the particles. $\omega_{eg}$ is the level-spacing between the excited $|e\rangle$ and ground $|g\rangle$ states of a single particle.

We observe that the amount of extracted work, one more time, is equal to the complete thermal energy of the condensate (system) at thermal equilibrium. This takes place again for a max entangled state, $|N,1\rangle$. We can take the excited state, e.g., as the motional states of a Bose-Einstein condensate with $\hbar\omega_{eg}=h^2/2mL^2$~\cite{moore1999quantum}. Then, the thermal energy of the condensate can be converted into the directional (kinetic) energy of a board placed in the condensate region. Here, again, the presented value of the extracted work is subject to the realization of the measured-particle in the excited state. The investigation of this symmetrization problem has further importance as entanglement of symmetrized many-body states and nonclassicality of photonic states are intimately related. A Dicke (many-body) state becomes a Fock (photon) state when $N\to \infty$~\cite{radcliffe1971some,klauder1985coherent,tasgin2017many}. Similarly, separable coherent atomic states become the photonic coherent states in the same limit.

\subsection*{Summary and Discussions}

{\it Summary.}--- We reinvestigate an already studied phenomenon --work extraction from an entangled system via measurement backaction~\cite{brunelli2017detecting,cuzminschi2021extractable}-- when nature itself performs the measurements. This is when there is no intellectual being~(such as a human, daemon, or an ancilla) is present for the measurement; but the monitoring is conducted by the environment/nature itself. In this case, measurement-basis becomes the coherent states~\cite{zurek1993coherent,gallis1996emergence,tegmark1994decoherence,wiseman1998maximally,zurek1994decoherence,zurek1995decoherence,paraoanu1999selection,zurek1993preferred}. This fixes the measurement strength to $\lambda=1$. The state of the $a$-mode, in which work-extraction is carried out, is already independent from the outcome of the $b$-mode~\cite{fiuravsek2002gaussian,giedke2002characterization,fiuravsek2007gaussian,giorda2010gaussian} as long as Gaussian states and measurements are concerned~\cite{fiuravsek2002gaussian,giedke2002characterization,fiuravsek2007gaussian,giorda2010gaussian}. (The measurement performed by the environment is a Gaussian one as the measurement-basis is coherent states.) Therefore, in total, the nature itself assigns a particular amount of work/energy to a given degree of entanglement.

Entanglement converts a disordered (randomly moving particles, microscopic) form of energy into an ordered form where the molecules of the board move along the same direction, i.e., macroscopic motion. The letter is referred as the mechanical energy, or we can tell that it is the KE.

We find that at low $T$, the directional energy associated with the entanglement is the ``total thermal energy'' times the degree of the entanglement for a two-mode Gaussian state: $W=\xi(r) U_{\rm ther}$. Here, $\xi (r) =\left[ 1-2/(1+\cosh 2r) \right]$ increases with the entanglement and gets closer to $\xi=1$ around the max entanglement. $r$ is the squeezing strength. That is, all thermal energy can be converted into directional energy for a maximally entangled Gaussian state. Similarly, the entire thermal energy can be converted into directional energy also for (ii) max entangled number state $(|0,1\rangle+|1,0\rangle)/\sqrt{2}$ and for (iii)~symmetrization entanglement of identical bosons~\cite{tasginSymmetrizationEnt}. However, the conversion in (ii) and (iii) is subject to the realization of the measured mode/particle in the excited state; while in (i) EIWE any measurement outcome extract that amount of work.

{\it Squeezing and potential energy}.--- In this section, we would like to introduce a correspondence also between potential mechanical energy and single-mode nonclassicality, SMNc, (e.g., squeezing). Entanglement and SMNc are two different types of nonclassicalities~(quantumness). The two not only can be converted into each other via passive optical elements, such as a beam splitter~(BS), but they also satisfy a conservation-like  relations~\cite{ge2015conservation,arkhipov2016nonclassicality,arkhipov2016nonclassicality,arkhipov2016interplay}. When a cavity mode is squeezed, it cannot be converted into work directly. This is because, squeezing keeps the entropy unchanged. However, the situation changes when the squeezed cavity field leaks out through the mirror(s). The interaction between the cavity and the output modes $\sum_{\bf k} (g_k \hat{b}_{\rm k}^\dagger \hat{a}+H.c.)$ is in the form of a BS interaction. Thus, the cavity and the output modes get entangled~\footnote{We note that such a study should employ the entanglement quantifications for wave-packets~\cite{tasgin2020nonclassicality} which calculates entanglement of the cavity mode with all of the vacuum modes. This can be performed using the input-output formalism developed for the wave-packets~\cite{inputputputWPs}.  }. Environmental monitoring on the output modes $\hat{b}_k$~\footnote{One may keep the quality of the optical cavity very high, e.g., $10^5$ Hz~\cite{thompson2008strong}, so that environment monitors the output modes in a much shorter decoherence time over the mass of particles present around.} makes the cavity extract work. We can also view the process as the potential mechanical energy~(associated with squeezing) transforms into the kinetic mechanical energy~(associated with entanglement).

{\it Connection with a recent study.}--- As a final but important point, we indicate that the present research actually investigates the results of a recent study~\cite{tasginEntanglementKKRs}. Rigorous calculations, employing the standard methods, clearly show an intriguing phenomenon in an optical cavity. Onset of entanglement exhibits itself in the response functions of the optical cavity. At this point, nonanalyticity of the response function moves into the upper-half of the complex frequency plane~(UH-CFP). One needs to avoid this incident from implying the violation of causality. Fortunately, surveys~\cite{hollowood2008refractive,hollowood2007causality,hollowood2008causal} show that a nonanalyticity in the UH-CFP does not imply the violation of causality if there exists a small curvature in the background. In the present study, the total curvature of the cavity ($a$-mode) increases as the disordered thermal energy is converted into ordered directional kinetic energy of the board. That is, it indeed increases with the amount of entanglement.

\nocite{serafini2003symplectic}

\subsection*{Acknowledgements}

We gratefully thank Vural G\"{o}kmen for the motivational support, Bayram Tekin for the scientific support, Wojciech H. Zurek for letting us know his influential work~\cite{zurek1993coherent} and Alessio Serafini for his support on the continuous-variable quantum information. We acknowledge the fund TUBITAK-1001 Grant No. 121F141.

%


\bibliography{bibliography}
\newpage

\section{\bf  SUPPLEMENTARY MATERIAL 
	\\ for
	\\{\it Environmental-Induced Work Extraction~(EIWE) }}

In this supplementary material~(SM), we first obtain the environmentally extracted in a two-mode~(TM) squeezed thermal~(Gaussian) state in Sec.~1. We show that it is in the form $W=\xi(r)\times(\bar{n}\hslash \omega)$ at low-temperatures~($T$) ---e.g., room temperature for optical modes. Here, $\xi(r)$, given in Eq.~(\ref{W_TM_sqz}), quantifies the strength of the entanglement. We use von Neumann entropy~($S_V$) in our calculations, in difference to Ref.~\cite{brunelli2017detecting}.

Second, in Sec.~2, we show that the same form for the work extraction, i.e., $W=\xi(r)\times(\bar{n}\hslash \omega)$, appears also for other TM Gaussian states.

 \subsection*{1. {EIWE for two-mode squeezed thermal state}}

Initially, before the measurement, both modes, $a$ and $b$, are in thermal equilibrium with an environment at temperature $T$. When one carries out a Gaussian measurement on the $b$-mode, the enropy of the $b$-mode reduces below the one for the thermal equilibrium. When the $a$-mode re-thermalizes with the environment it performs a work in the amount of \cite{maruyama2009colloquium}
\begin{eqnarray}
W=k_B T (S_V^{(\rm ther)}-S_V^{(\rm meas)}).
\end{eqnarray}
Here, $S_V^{(\rm meas)}$ is the reduced entropy of the $a$-mode after the measurement in the $b$-mode is carried out. $S_V^{(\rm ther)}$ is the entropy of the $a$-mode after the rethermalization of the mode. 

Entropy of a Gaussian state can be determined by its covariance matrix, which includes the noise elements of the modes. Covariance matrix of a biparitite Gaussian state can be cast in the form~\cite{duan2000inseparability}
\begin{eqnarray}
\sigma_{ab}= 
\begin{bmatrix} 
   \sigma_{a} &  c_{ab}\\
    c_{ab}^T  &   \sigma_{b} 
\end{bmatrix}
\end{eqnarray}
via local symplectic transformation $Sp(2,\mathbb{R})\oplus Sp(2,\mathbb{R})$-i.e., transformations altering neither of the entropy or entanglement. Here, $\sigma_{a}={\rm diag}(a,a)$ and  $\sigma_{b}={\rm diag}(b,b)$ are the reduced covariance matrices of the $a$ and $b$ modes, respectively. $c_{ab}={\rm diag}(c_1,c_2)$ refers to correlations/entanglement between the two mode. For a symmetrically squeezed two-mode thermal state, i.e., both modes used to be in thermal equilibrium with the same $T$ in the squeezing process, $b=a$ and $c_2=-c_1=-c$.

The state into which the $a$-mode collapses is independent from the outcome of the $b$-mode measurement as long as a Gaussian measurement is carried out~\cite{fiuravsek2002gaussian,giedke2002characterization,fiuravsek2007gaussian,giorda2010gaussian}. The covariance matrix of the $a$-mode after the measurement becomes
\begin{eqnarray}
\sigma_{a}^{\pi_b}= \sigma_a-c_{ab}\:(\sigma_b+\gamma^{\pi_b})^{-1} \: c_{ab}^T.
\end{eqnarray}
Here, $\gamma^{\pi_b}=R(\phi) \: {\rm diag}(\lambda/2, \lambda^{-1}/2)\: R(\phi)^T$ refers to the covariance matrix associated with a Gaussian operation (measurment)~\cite{fiuravsek2002gaussian,giedke2002characterization,fiuravsek2007gaussian,giorda2010gaussian}. $\lambda$ is the measurement strength. For a Gaussian measurement having the coherent states as a basis, $\lambda=1$ and $\gamma^{\pi_b}= {\rm diag}(1/2, {1}/2)$ independent of the rotations $R(\phi)$ in the $a$-mode, i.e., $\hat{a}\rightarrow \hat{a} e^{i \phi}$.

The entropy of a Gaussian state is determined solely by purity, $\mu=\frac{1}{2^n \sqrt{{\rm Det} \sigma}}$, which takes the form
\begin{eqnarray}
S_V= \frac{1-\mu}{2\mu} \ln \left(\frac{1+\mu}{1-\mu}\right)-\ln \left(\frac{2\mu}{1+\mu}\right)
\end{eqnarray}
for a single-mode state~\footnote{Please note that here we use von Neumann entropy in difference to Ref.~\cite{brunelli2017detecting}, where Renyi entropy is employed.}.

The purity of the $a$-mode, after the $b$-mode measurement, can be obtained as 
\begin{eqnarray}
\mu_1\equiv\mu^{\rm (meas)}= \frac{a+1/2}{2(a^2-c^2+a/2)}.
\end{eqnarray}
For a TM squeezed thermal state,
\begin{eqnarray}
a&=&(\bar{n}+\frac{1}{2})\cosh(2r),\\
c&=&(\bar{n}+\frac{1}{2})\sinh(2r),
\end{eqnarray}
the purity becomes 
\begin{eqnarray}
\mu_1=\frac{a+1/2}{2(\bar{n}+1/2)^2+a},
\end{eqnarray}
where $\bar{n}=(e^{\hbar \omega_a/k_BT}-1)^{-1}$ is the occupation of the $a$-mode, which becomes $\bar{n}\rightarrow e^{-\hbar \omega_a/k_BT}$ at low T--- e.g., the room temperature for optical modes of resonance $\omega_a$. $r$ is the two-mode squeezing strength with which entanglement increases~\cite{ScullyZubairyBook}.  

$\bar{n}$ is extremely small at low T regime. So, it is
\begin{eqnarray}
\mu_1\cong 1-\frac{2\bar{n}}{a+1/2}.
\end{eqnarray} 
Then, the entropy can be approximately written as,
\begin{eqnarray}
S_V^{(\rm meas)}\cong &&\frac{\bar{n}}{a+1/2}[\ln(2)-\ln(2\bar{n})+\ln(a+\frac{1}{2})] \qquad \\ \nonumber
&&- \frac{2\bar{n}}{a+1/2} ,
\end{eqnarray} 
where $a=(\bar{n}+1/2)\cosh(2r)$. The last term originates from the $\ln \left(\frac{2\mu}{1+\mu}\right)$ term given in Eq.~(S4).

In the square brackets, in Eq.~(S10), $\ln(\bar{n})=\frac{\hbar \omega_a}{k_BT}\gg 1$ and $\ln(a+1/2)\cong \ln(\cosh(2r))$. Assuming that the squuezing rate is much smaller than $\frac{\hbar \omega_a}{k_BT}$, which is about $\sim$100 at the room temperatures, i.e., $r\ll\frac{\hbar \omega_a}{k_BT}$, the entropy takes the form
\begin{eqnarray}
S_V^{(\rm meas)}\cong \frac{2\bar{n}}{1+\cosh(2r)}\frac{\hbar \omega_a}{k_BT},
\end{eqnarray} 
where the last term in Eq.~(S10) is also neglected. 

Some time after the measurement, $a$-mode rethermalizes with the enviroment and at equilibrium its purity becomes
\begin{eqnarray}
\mu_2 \equiv \mu^{(\rm ther)}=\frac{1}{1+2\bar{n}}.
\end{eqnarray} 
The entropy at equilibrium can similarly calculated as 
\begin{eqnarray}
S_V^{(\rm ther)}\cong \bar{n}\frac{\hbar \omega_a}{k_BT}.
\end{eqnarray} 
Therefore, the extracted work becomes
\begin{eqnarray}
W=\left(1-\frac{2}{1+\cosh(2r)}\right) \bar{n}\hbar \omega_a,
\label{W_TM_sqz}
\end{eqnarray}
where the $k_BT$ term in Eq.~(S1) is canceled by $ 1/ k_BT$ appearing in (S11) and (S13).

\subsection*{2. EIWE for other Gaussian states} 

We derived the simple form for the extracted work $W=\xi(r)(\bar{n}\hbar \omega_a)$ for the TM squeezed thermal states. Now, we aim to show that a similar form appears also for other Gassian states given by the covariance matrix~(S2).

When the $b$-mode is measured by the enviroment, the $a$-mode collapses to a covariance matrix with determinant
\begin{eqnarray}
{\rm det} \sigma_a^{\pi_b}=\frac{(a^2-c_1^2+a/2)(a^2-c_2^2+a/2)}{(a+1/2)^2}.
\end{eqnarray}
The entropy after the measurement is similarly $\mu^{(\rm meas)}=\frac{1}{2\sqrt{{\rm det} \sigma_a^{\pi_b}}}$~\cite{serafini2003symplectic}. For $c_1=-c_2=c$, $\mu^{(\rm meas)}$ becomes the purity given in Eq.~(S5).

As we aim to show that the extracted work has a ``form'' similar to $W=\xi\cdot(\bar{n}\hbar \omega_a)$ also for other Gaussian states, we express Eq.~(S15) in terms of $Sp(4,\mathbb{R})$ invariants
\begin{eqnarray}
{\rm det}\sigma &=&(a^2-c_1^2)(a^2-c_2^2), \\
\Delta &=& 2(a^2+c_1c_2),
\end{eqnarray}
where ${\rm det} \sigma$ is the determinant of the two-mode state before the measurement. We do this because any one of the two-mode Gaussian states, Eq.~(S2), can be obtained from $Sp(4,\mathbb{R})$ transformations of the TM squeezed thermal state~Eq.~(15) of Ref.~\cite{serafini2003symplectic}. The determinant in Eq.~(S15) can be expressed as 
\begin{small}
\begin{eqnarray}
{\rm det}\sigma^{\pi_b}_a =\frac{(a^2-c_1^2)(a^2-c_2^2)+\frac{a}{2}(2a^2-c_1^2-c_2^2)}{(\frac{1}{2}+a)^2}, 
\end{eqnarray}\end{small}
where the first term is the $Sp(4,\mathbb{R})$ invariant ${\rm det}\sigma$. In the second term, $I=2a^2-c_1^2-c_2^2$ can be expressed in terms of $Sp(4,\mathbb{R})$ invariants~(Please note that $a$ is only local $Sp(2,\mathbb{R})$ invariant) as
\begin{eqnarray}
{\rm det} \sigma^{\pi_b} =\frac{(a^2 I+ \Delta^2/4-\Delta a^2)}{(a+\frac{1}{2})^2}.
\end{eqnarray}
We note that ${\rm det} \sigma=(\tilde{a}^2_{\rm TMS}-\tilde{c}^2)^2$ and $\Delta=2(\tilde{a}^2-\tilde{c}^2)$ for the TM squeezed thermal states and they are $Sp(4,\mathbb{R})$ invariant. Thus, Eq.~(S19) can be recast as
\begin{eqnarray}
(\tilde{a}^2-\tilde{c}^2)^2=a^2I+(\tilde{a}^2-\tilde{c}^2)^2-\Delta a^2.
\end{eqnarray}
Please note that, in this section, we use tilde symbol for the covariance matrix elements of the TM squeezed thermal states in order to distinguish them from the variable $a$ given in the general matrix given in Eq.~(S2).

Cancellation in Eq.~(S20) results
 \begin{eqnarray}
I=2a^2-c_1^2-c_2=\Delta.
\end{eqnarray}
Using this in Eq.~(S19), we obtain the expression
 \begin{eqnarray}
\mu^{(\rm meas)}=\frac{a+1/2}{2(z+a/2)},
\end{eqnarray}
where $z=\tilde{a}^2-\tilde{c}^2=(\bar{n}+1/2)^2$. Please note that Eq.~(S22) is in the same form with Eq.~(S5) from which we obtain the extracted work 
 \begin{eqnarray}
W=\xi \: (\bar{n}\hbar \omega_c).
\end{eqnarray}
Thus, above we showed that Eq.~(S23) is a general form for the extracted work from a symmetric TM Gaussian state.


\end{document}